\documentclass{article}

\usepackage{amsfonts, amssymb}
\usepackage{amsmath}
\usepackage{amsthm}
\usepackage{color}

\newtheorem{theorem}{Theorem}
\newtheorem{definition}{Definition}
\newtheorem{proposition}{Proposition}
\newtheorem{lemma}{Lemma}

\def\om{{\omega}}
\def\db#1{\mathbb #1}
\def\cal#1{{\mathcal #1}}
\def\esm#1{{\db E\left[\,#1\, \right]}}
\def\pr#1{{\db P\left\{\,#1\,\right\}}}

\def\QED{{$_\blacksquare$}}

\def\eps{\epsilon}
\def\one{{\bf 1}}

\def\psn{\par\smallskip\noindent}
\def\pmn{\par\medskip\noindent}

\def\Lam{\Lambda}

\def\cB{{\cal B}}
\def\cC{{\cal C}}

\def\cF{{\cal F}}

\def\cH{{\cal H}}

\def\cT{{\cal T}}

\def\uC{{\mathbf C}}
\def\uH{{\mathbf H}}

\def\uu{{\mathbf u}}

\def\BLam{{\mathbf \Lambda}}

\def\om{\omega}
\def\Om{\Omega}
\def\Lam{\Lambda}
\def\th{\theta}
\def\Th{\Theta}
\def\ffi{\varphi}

\def\dist{{\rm{dist}}}
\def\one{{\mathbf 1}}

\def\dist{{{\rm dist }}}

\def\diam{{{\rm diam\, }}}
\def\card{{{\rm card\, }}}
\def\supp{{{\rm supp\, }}}
\def\myset#1{{\left\{\, #1 \, \right\}}}

\def\myset#1{{\left\{\,#1\,\right\}}}

\def\P{\mathbb P}
\def\Q{\mathbb Q}
\def\R{\mathbb R}
\def\T{\mathbb T}
\def\Z{\mathbb Z}

\usepackage{latexsym}
\usepackage{amsfonts, amssymb}
\usepackage{amsmath}
\usepackage{graphicx}
\usepackage{array}
\usepackage{color}
\usepackage{stmaryrd}

\numberwithin{equation}{section}

\begin{document}

\title{Wegner-type bounds for a two-particle lattice model with a generic "rough" quasi-periodic potential }

%\subtitle{Do you have a subtitle?\\ If so, write it here}
\author{Martin Gaume % etc
}                     % Do not remove
\date{}
% The correct dates will be entered by Springer

\maketitle
\pmn
Universit\'{e} Paris Diderot - IMJ \\
Site Chevaleret\\
case 7012\\
5  rue Thomas Mann\\
75205 Paris cedex 13\\
gaume@math.jussieu.fr\pmn 
\pmn

\begin{abstract}
In this paper, we consider a class of two-particle tight-binding Hamiltonians, describing pairs of interacting quantum particles on the lattice $\Z^d$, $d\ge 1$,  subject to a common external potential $V(x)$ which we assume quasi-periodic and depending on auxiliary parameters.  Such parametric families of ergodic deterministic potentials ("grand ensembles") have been introduced earlier in \cite{C07}, in the framework of single-particle lattice systems, where it was proved that a non-uniform analog of the Wegner bound holds true for a class of quasi-periodic grand ensembles. Using the approach
proposed in \cite{CS08}, we establish volume-dependent Wegner-type bounds for a class of quasi-periodic two-particle lattice systems with a non-random short-range interaction.
\end{abstract}
\section{Introduction}\label{intro}

$$
H(\om;\th) = \sum_{j=1}^2 \left(\Delta_j + V(x_j;\om;\th) \right) + U(x_1,x_2),
\eqno(1.1)
$$
where
$$
\begin{array}{l}
(\Delta_1 f) (x_1, x_2) = f(x_1 - 1, x_2) + f(x_1 + 1, x_2), \\
(\Delta_2 f) (x_1, x_2) = f(x_1, x_2 - 1) + f(x_1, x_2 + 1).
\end{array}
$$
The function $V:\Z^d\times \T^\nu \times \Th \to \R$ is defined as follows.
For every $n\ge 1$, consider the partition of the unit circle $\T^1 = \R^1/\Z^1$ into intervals
$I_{n,i}=\left[\frac{i-1}{2^n}, \frac{i}{2^n} \right)$ of size $2^{-n}$, $i=1, \ldots, 2^n$. Let
$\cC_n = \left\{\uC_{n,k}, k=1, \ldots, 2^{\nu n}\right\}$ be the family of all possible Cartesian products
$I_{n,i_1} \times \cdots \times I_{n,i_\nu}$, i.e. a partition of the torus $\T^\nu$ into a family of cubes of sidelength $2^{-n}$. Further, let $\ffi_{n,k} = \one_{\uC_{n,k}}$ be the indicator functions of the cube $\uC_{n,k}$. Next, introduce a parametric family of functions on the torus
$$
v(\om;\th) = \sum_{n=1}^\infty a_n \, \sum_{k=1}^{K_n} \th_{n,k} \ffi_{n,k}(\om),
\eqno(1.2)
$$
where $\{\th_{n,k}, n\ge 1, 1 \le k \le K_n < \infty\}$ are IID random variables on some auxiliary probability space $\Th$, with uniform distribution in $[0,1]$. We will identify $\Th$ with the set of all samples
$\{\th_{n,k}, n\ge 1, 1 \le k \le K_n\}$. Let $\{T^x, x\in\Z^d\}$ be an action of the additive group $\Z^d$ on the torus $\T^\nu$, which we assume to satisfy a Diophantine condition of the form
$$
\forall\, x\in\Z^d\;\; \dist(\om, T^x \om) \ge \frac{Const}{ \|x\|^B}, \; 0 < B < \infty.
\eqno(1.3)
$$
Concerning the rate of decay of the coefficients $a_n$ in the expansion (1.2), we assume that
$\forall\, n\ge 1$,
$$
  \frac{c''}{n^{M}} \le |a_n| \le\frac{c'}{n^{\kappa}}
\eqno(1.4)
$$
for some $c',c''\in(0,+\infty)$, $1 < \kappa \le M < \infty$. Notice that the upper bound
on $|a_n|$ guarantees the convergence of the above expansion, while the lower bound is required for our method.

Finally, we set
$$
V(x;\om;\th) = v(T^x \om; \th), \; x\in\Z^d, \; \om\in\Om \equiv\T^\nu, \; \th\in\Th.
\eqno(1.5)
$$
Following \cite{C07}, we will call such a parametric family of functions on $\Om$ a grand ensemble of randelette type.

The expansions of the form (1.2) will be called  randelette expansions.

\section{Wegner-type bounds. Main results}

%\section{Grand ensembles of deterministic potentials}

In the spectral theory of random operators, e.g. Anderson tight-binding Hamiltonians of the form
$$
H(\om) = \Delta + V(x;\om), \; x\in\Z^d,
\eqno(2.1)
$$
with an IID random potential $V(x;\om)$, an important role is played by eigenvalue concentration bounds. The first fairly general result of such kind was obtained by F. Wegner \cite{W81}, so they are usually called Wegner-type bounds. Namely, consider the lattice Hamiltonian of the form (2.1), where the random variables $V(x;\cdot)$, identically distributed with a common probability cumulative distribution function (CDF, for short) $F(s) = F_V(s)$, admit a bounded probability density $p(s) = p_V(s)$, so that
$$
\| p_V \|_\infty < \infty,
\eqno(2.2)
$$
Given a finite lattice cube $\Lam_L(u)\subset\Z^d$ of an arbitrary center $u\in\Z^d$ and of sidelength $2L+1, L\ge 0$, consider the restriction of the Hamiltonian (???) on $\Lam \equiv \Lam_L(u)$ with Dirichlet boundary conditions. Further, let
$$
\Sigma(H_{\Lam}(\om)) = \{E_j^\Lam, \, j=1, \ldots, |\Lam| = (2L+1)^d \}
$$
be the (random) spectrum of the finite-volume Hamiltonian $H_{\Lam}(\om)$, i.e. the set of its (random) EVs counted with multiplicities. Then for all $E\in \R$ and all $\eps\in(0,1)$, we have
$$
\begin{array}{l}
 \;\;\;\;\pr{ \dist\left(\Sigma(H_{\Lam}(\om)), E\right) \le \eps } \\
 = \pr{ \exists E^\Lam_j \in [E-\eps, E+\eps] }\\
  \le |\Lam| \cdot \| p_V \|_\infty \cdot \eps.
 \end{array}
 \eqno(2.3)
$$
The presence of the factor $|\Lam| \equiv \card \Lam$ at the RHS of (2.3) allows to prove the absolute continuity of the so-called (limiting) density of states
$$
N(E) = \lim_{L\to\infty} \frac{1}{|\Lam_L(0)|} \card\left\{j: \; E_j^\Lam\le E \right\},
$$
where the limit exists with probability one and is non-random (cf. \cite{CyFKS87}, \cite{CarL90}, \cite{PF92}  and references therein). In other words, the absolute continuity of the marginal CDF $F_V(s)$ implies that $N(E)$ also admits a density, called density of states (DoS) $\rho(E)$, so that
$$
N(E) = \int_{-\infty}^E \rho(E') \, dE'.
$$

Many generalizations of the Wegner bound (2.3) are well-known by now. In particular, analogs of the Wegner bound can be given in cases where the marginal distribution of the random potential field $V(x;\om)$ does not admit a density, but the CDF $F_V(s)$ is H\"{o}lder-continuous; see, e.g., \cite{St00, St01}. On the other hand, in the case of sufficiently regular marginal distributions optimal eigenvalue concentration bounds are known; see \cite{CoHK07}.

The main difficulty for an extension of the Wegner bound to deterministic (e.g., quasi-periodic) potentials is that all known "probabilistic" methods applicable to the EV concentration problem require a greter freedom in varying individual values of the potential. For example, the conventional Wegner bound requires the RV $V(x;\cdot)$ to be independent. It is well-understood by now that the independence requirement can be replaced by that of asymptotic decay of dependence. In \cite{CS08}, even a greater degree of correlation was allowed. Still, an ensemble of quasi-periodic potentials
$$
V(x;\om) = v(\om + x\alpha),  \; \om\in\T^1 = \R^1/\Z^1, \; \alpha\in\R\setminus \Q,
$$
is too "rigid" and  does not allow a direct application of probabilistic methods to the EV concentration problem.

A reader interested in eigenvalue concentration bounds and,  in particular, in the analysis of regularity of the density of states, can find an extensive bibliography in references \cite{BCKP88}, \cite{CamK86}, \cite{CoHK07}, \cite{ConFS83}, \cite{CrSim83}, \cite{ST85}, as well in the above mentioned monographs \cite{CyFKS87}, \cite{CarL90},
\cite{PF92}, \cite{St01}.

In this connection, we mention an alternative powerful approach developed earlier by J. Bourgain, M. Goldstein and W. Schlag (see, e.g., \cite{BG00, BS00, BGS01}). Unfortunately, their approach requires the function $v:\T^\nu\to\R$ to be \textit{analytic}, which greatly limits the applications.

This is why we use in the present paper an approach proposed in \cite{C07}, where, instead of an individual ergodic ensemble of quasi-periodic potentials $\{V(\cdot;\om), \om\in\T^\nu\}$ a parametric family
$\{V(\cdot;\om;\th), \om\in\T^\nu, \th\in\Th\}$ labeled by points of a specially constructed parameter space $\Th$ is considered. While such an ensemble, (called "grand ensemble" in \cite{C07}) is not ergodic, it turns out that
some analog of Wegner-type bounds can be established for \textit{generic} parameter values $\th\in\Th$.

A drawback of this method is that obtained EV concentration bounds are non-uniform, unlike the conventional Wegner-type bounds.

The main results of the resent paper are given in the following two statements. For notational simplicity, below we will denote by $\Sigma_{\uu,L}(\om;\th))$ the spectrum (i.e., the set of eigenvalues counted with multiplicities) of operator $\uH_{\BLam_L(\uu)}(\om;\th))$:
$$
\Sigma_{\uu,L}(\om;\th)) = \uH_{\BLam_L(\uu)}(\om;\th))
= \left\{ E^{\Lam_L(\uu)}_j, \; 1 \le j \le |\Lam_L(\uu)| \right\}.
\eqno(2.4)
$$
\begin{theorem}\label{ThmMain1}
Consider an ensemble of functions on the lattice $V_\th(x;\om) = v(T^x \om; \th)$, $x\in\Z^d$, of the form (1.2) labeled by points of the torus $\T^\nu$ and by parameter $\th\in\Th$. Fix two numbers,  $b>0$ and $r>1$. Given a lattice cube $\BLam_L(\uu)\subset\Z^{2d}$, $\BLam_L(\uu)\subset \Lam_{L^r}(v)$, there exists a subset $\Th_L\subset\Th$ of measure $\mu(\Th_L) \ge 1 - L^{-b}$, $b>0$, such that for any $\th\in\Th_L$ the following inequality holds (with $\th\in\Th_L$ fixed):
$$
\pr{\om\in\T^\nu:\, \dist[ \Sigma_{\uu}(\om;\th)), E]\le \eps }
\le Const\, L^{M+b+3d+r} \, \eps.
$$
\end{theorem}

\begin{theorem}\label{ThmMain2} Under the same assumptions as in Theorem \ref{ThmMain1}, consider two lattice cubes
$\BLam_L(\uu')$, $\BLam_L(\uu'')\subset \Z^{2d}$, such that
$$
\|\uu' - \uu''\|\le L^{r}, \, r > 1,
$$
and
$$
 \min \{ \|\uu -\uu'\|, \|\uu -S(\uu')\| \} > 8 L
 \eqno(2.5)
$$
where  $S:(u_1,u_2)\mapsto (u_2,,u_1)$ is the symmetry in $\Z^{2d}$ exchanging the coordinates of two particles.
Given a number $b\in(0,+\infty)$, there exists a subset $\Th_L\subset\Th$ of measure $\mu(\Th_L) \ge 1 - L^{-b}$, $b>0$, such that for any $\th\in\Th_L$ the following inequality holds (with $\th\in\Th_L$ fixed):
$$
\pr{\om\in\T^\nu:\, \dist[ \Sigma_{\uu'}(\om;\th)), \Sigma_{\uu''}(\om;\th)) ]
\le \eps } \le Const\, L^{M+b+r+5d} \, \eps.
$$
\end{theorem}

The proofs of theorems \ref{ThmMain1} and \ref{ThmMain2} are given in Subsections \ref{SubSectQPW1} and \ref{SubSectQPW2}, respectively.

\pmn
\textbf{Remark.} To explain the role of condition (2.5), recall that the potential energy of a two-particle system is invariant under the symmetry $S:\,(u_1, u_2) \mapsto (u_2,u_1)$. Given a two-particle volume
$\BLam_L(u_1,u_2) = \Lam_L(u_1) \times \Lam_L(u_2)$, consider its "total shadow",
$$
\Pi \BLam_L(u_1,u_2) :=  \Lam_L(u_1) \cup \Lam_L(u_2).
$$
Obviously, $\Pi \BLam_L(u_1,u_2) = \Pi \BLam_L(u_2,u_1)$, i.e.,
$$
\Pi \BLam_L(\uu) = \Pi \BLam_L( S(\uu)) = S(\BLam_L(\uu)),
$$
so that the samples  $\{V(x;\om), x \in \Pi \BLam_L(\uu)\}$ and
$\{V(x;\om), x \in \Pi \BLam_L(S(\uu) )\}$ are identical. As a consequence, Hamiltonians $\uH(\BLam_L(\uu))$
and $\uH(\BLam_L(S(\uu) ))$ have identical spectra. Further discussion of this condition can be found in \cite{CS08}.
\pmn

Naturally, the above upper bounds are useful only for sufficiently small $\eps$, when the RHS is smaller than $1$.

\section{Diagonally monotone operator families}

Here we briefly recall some notions and results from \cite{St00}, \cite{St01} and their extensions to two-particle systems proposed in \cite{CS08}.

  Let $m\geq 1$ be a positive integer, and $J$ an abstract finite set with $|J| (=\card J) = m$.
Consider the Euclidean space $\db R^J \cong \R^m$ with standard basis
$(e_1, \dots, e_m)$, and its positive orthant
$$
\db R^J_+ = \myset{ q\in\db R^J:\, q_j\geq 0, \,\, j=1, 2, \dots, m }.
$$
For any measure $\mu$ on $\db R$, we will denote by $\mu^m$ the product measure $\mu
\times \dots \times \mu$ on $\db R^J$. Furthermore, for any probability measure $\mu$ and
for any $\eps>0$, define the following quantity:
$$
s(\mu, \eps) = \sup_{a\in\db R} \,\, \mu([a, a+\eps])
$$
Furthermore, let $\mu^{m-1}$ be the marginal
probability distribution induced by $\mu^m$ on $q'=(q_2, \dots, q_m)$.
\begin{definition} Let $J$ be a finite set with $\,|J| = m$.
Consider a function $\Phi:\, \db R^J\to\db R$. It is called diagonally monotone (DM, for short) if it satisfies the following conditions:
\psn
{\rm (1)} for any $r\in\db R^J_+$ and any $q\in \db R^J$,
$$
\Phi(q+r)\geq \Phi(q);
$$
\psn
{\rm(2)} moreover, for $e=e_1 + \dots + e_m\in \db R^J$, for any $q\in\db R^J$
and for any $t>0$
$$
\Phi(q + t \cdot e) - \Phi(q) \geq t.
$$
\end{definition}

  It is convenient to introduce the notion of DM operators considered
as  quadratic forms. In the following definition, we use the same notations as above.

\begin{definition} Let $\cH$ be a Hilbert space. A family of self-adjoint operators
$B(q): \cH \to \cH$, $q\in \R^J$, is called DM if,
$$
\forall\,q\in\R^J\,\; \forall\, r\in\R^J_+
\;\; B(q+r) \geq B(q),
$$
in the sense of quadratic forms, and for any vector
$f\in\cH$ with $\|f\|=1$, the function $\Phi_f: \R^J \to \R$ defined by
$$
\Phi_f(q) = (B(q)f, f)
$$
is DM.
\end{definition}

\pmn
\textbf{Remark. } By virtue of the variational principle for self-adjoint operators,
if an operator  family $H(q)$ in a finite-dimensional Hilbert space $\cH$ is
DM, then each eigen-value $E_k^{B(q)}$ of $B(q)$ is a DM function. In addition, it is readily seen
that if  $H(q), \, q\in\R^J$,  is a DM  operator  family
in Hilbert space $\cH$, and $H_0:\cH\to\cH$ is an arbitrary self-adjoint operator, then
the family $H_0 + H(q)$ is also DM.
\pmn

\begin{lemma}[Stollmann, \cite{St00}]\label{LemStollmann} Let $J$ be a finite index set,
$\mu$ be a probability measure on $\db R$, and $\mu^J$ be the product measure on $\R^J$
with marginal measures $\mu$. If the function $\Phi:\, \db R^J \to \db R$ is
DM, then for any open interval $I\subset \db R$ we have
$$
\mu^J\myset{ q:\, \Phi(q) \in I } \leq |J| \cdot s(\mu, |I|).
$$
\end{lemma}

\textbf{Remark. } It is not difficult to see that the identical distribution of the RV $V$ is not essential for the Stollmann's bound. In a more general case, when the measure $\mu^J$ is a direct product
 of measures $\{\mu_j, j\in J\}$,  the above bound can be replaced by
$$
\pr{ q:\, \Phi(q) \in I } \le |J| \cdot|\Lam|^2 \cdot \max_{j} s(\mu_j, I).
$$
In the particular case where $\mu_j$ admit a bounded probability densities $p_j(t)$, the Stollmann's bound takes the form
$$
\pr{ q:\, \Phi(q) \in [a,a+\eps] } \le |J| \cdot|\Lam|^2 \cdot \max_{j} \|p_j\| \, \eps.
$$

The proof of Stollmann's lemma can be found in \cite{St00}, so we omit it here.
\pmn

As was shown in \cite{CS08}, an ensemble of two-particle tight-binding Hamiltonians
$\uH_{\BLam} = \uH_{0,\Lam} + U + (V(x_1;\om) + V(x_1;\om))$ in a finite cube
$\BLam = \Lam^{(1)} \times \Lam^{(2)}\subset \Z^{2d}$ with external potential $V(x;\om)$ taking independent values is a diagonally monotone family, relative to the index set $J = \Pi \BLam := \Lam^{(1)} \cup \Lam^{(2)}$. This allows a fairly straightforward aplication of Stollmann's bound to two-particle systems. In our case, we achieve a similar goal, using a particular conditioning in the "grand ensemble" of quasi-periodic potentials $V(x;\om;\th)$; see Subsection
\ref{SubSectCondInd}.

\section{Proof of the main result}\label{SectProofs}
%\bigskip

\subsection{Spacing along finite-volume trajectories}

Owing to the Diophantine condition (1.3), for any cube $\Lam_{L}(u)\subset\Z^d$,
$L\ge 1$, and any point  $\om\in\T^\nu$, we have
$$
\delta(\Lam_L(u),\om) := \min_{x,y\in\Lam_L(u), \, x\neq y }
\left\{ \dist(T^x \om, T^y \om) \right\} \ge C L^{-B}
\eqno(4.1)
$$
for some $C>0$. In fact, $\delta(\om)$ does not depend on $\om$, since the conventional distance on the torus is shift-invariant, so that below we drop the argument $\om$ in $\delta(\Lam_L(u))=\delta(\Lam_L(u),\om)$. For the same reason, $\delta(\Lam_L(u))$ does not depend upon $u$, and we will use a simpler notation $\delta_L$ for the quantity $\delta(\Lam_L(u))$.

Given a positive integer $n$, consider again the partition $\cC_n$ of the torus $\T^\nu$ into cubes $\uC_{n,k}$,
$k=1, \ldots, 2^{\nu n}$, of the form $\uC_{n,k}=I_{n,i_1} \times \cdots \times I_{n,i_\nu}$, with
$I_{n,i}=\left[\frac{i-1}{2^n}, \frac{i}{2^n} \right)$, as defined in Section 1. It is convenient for our purposes to use the $\max$-norm in $\R^\nu$,
$$
\| \om \|_\infty := \max_{1 \le i \le \nu} |\om_j|, \; \om\in\R^\nu,
$$
and the distance induced by it on $\T^\nu = \R^\nu/\Z^\nu$. Below we always use this distance on the torus, unless otherwise specified. Then $\diam \uC_{n,k} = 2^{-n \nu}$, and we see that points of any finite-volume trajectory
$$
\cT(\om, \Lam_L(u)) = \{T^x \om, \, x\in\Lam_L(u)\}
$$
are separated by elements of any partition $\cC_n$ with
$$
n \ge n_0(L) = \ln \delta_L / \ln 2 \ge C' \ln L
\eqno(4.2)
$$
for some $0 < C' < \infty$. As a consequence, given any point $\om\in\T^\nu$, a positive integer $n_0$ and a lattice cube $\Lam_L(u)$, a finite family of RVs
$$
\left\{ \th_{n_0,k}: \, \supp \ffi_{n_0,k} \cap \cT(\om, \Lam_{L}(u)) \neq \varnothing \right\}
\eqno(4.3)
$$
(with $n_0$ fixed) is \textit{independent}. Recall also that each of the above RVs $\th_{n_0,k}$ is uniformly distributed in $[0,1]$, so that its probability density is bounded by $1$.
%%% %%%
Moreover , for any lattice cube $\BLam_L(\uu)\subset\Z^{2d}$ , so that $\Pi \BLam_L(\uu)\subset\BLam_N(v)$,we see that points of any finite-volume trajectory
$$
\cT(\om, \Pi\BLam_L(\uu)) = \{T^x \om, \, x\in\Pi\BLam_L(\uu)\}
$$
are separated by elements of any partition $\cC_n$ with
$$
n \ge n_0(N)
$$

\subsection{Conditional independence of the potential values}\label{SubSectCondInd}

Now we analyze the values of the function $v:\, \T^\nu\times \Th \to \R$ along the points of a given finite trajectory $\cT(\om, \Pi\BLam_{L}(\uu))$. Relative to the product measure $d\om \times \mu$ on the product probability space
$\T^\nu\times \Th$, these values are \textit{not} independent. (Here, $d\om$ is the normalized Haar measure on the torus.)  However, they are \textit{conditionally} independent given the sigma-algebra $\cB(\Lam_N(v))$ generated by the RVs $\om_i, 1\le i \le \nu$ and by all RVs $\{\th_{n,k}:\, n < n_0(N)\}$. Indeed, we can re-write the expansion (1.2)
("randelette" expansion) as follows:
$$
\begin{array}{l}
v(\om;\th) = \displaystyle\sum_{n=1}^\infty a_n \, \sum_{k=1}^{K_n} \th_{n,k} \ffi_{n,k}(\om) \\
= \displaystyle\sum_{n < n_0(N)} a_n \, \sum_{k=1}^{K_n} \th_{n,k} \ffi_{n,k}(\om)
+ \sum_{n \ge n_0(N)} a_n \, \sum_{k=1}^{K_n} \th_{n,k} \ffi_{n,k}(\om) \\
\end{array}
\eqno(4.3)
$$
It is straightforward now that the first sum at the RHS becomes constant, given the sigma-algebra $\cB(\Lam_N(v))$.
Fix two points $\om', \om''\in\T^\nu$ with $\dist(\om', \om'') \ge \delta_N$. For any $n\ge n_0(\Lam_L(u)$, they are separated by the elements of partition $\cC_n$. Observe that, actually,
$$
 a_n \, \sum_{k=1}^{K_n} \th_{n,k} \ffi_{n,k}(\om')
= a_n \,  \th_{n,k(\om')} \ffi_{n,k(\om')}(\om')
$$
where $k(\om')$ is uniquely defined by the condition
$$
\om' \in \supp \ffi_{n_0,k} \eqno(4.4)
$$
Further, with $k(\om'')$ defined in a similar way for the point $\om''$, $k(\om') \neq k(\om'')$, since $\om'$ and $\om''$ are separated by elements of $\cC_n$. This yields the representations
$$
\begin{array}{l}
v(\om';\th) = \xi' + \displaystyle\sum_{n \ge n_0(N)}^\infty a_n \,  \th_{n,k(\om')} \ffi_{n,k(\om')}(\om') = \xi' +\eta', \\
 v(\om'';\th) = \xi'' + \displaystyle\sum_{n \ge n_0(N)}^\infty a_n \,  \th_{n,k(\om'')} \ffi_{n,k(\om'')}(\om'') = \xi'' +\eta'',
\end{array}
\eqno(4.5)
$$
where $\xi', \xi''$ are $\cB(\Lam_N(v))$-measurable and $\eta', \eta''$ are conditionally independent, given $\cB(\Lam_N(v))$. Actually, the entire family of the RVs $v(T^x\om;\th)$, $x\in\Pi\BLam_L(\uu)$, admits the  decomposition
$$
\begin{array}{ll}
v(T^x\om;\th)  &= \xi_x +  \displaystyle\sum_{n \ge n_0(N)}^\infty  a_n \,  \th_{n,k(T^x\om)} \ffi_{n,k(T^x\om)}(T^x\om)\\
 &=  \xi_x +\eta_x,
\end{array}
\eqno(4.6)
$$
where all $\xi_x$ are $\cB(\Lam_N(v))$-measurable and the family of RVs $\{\eta_x, \, x\in\Lam_L(u)\}$ is independent. Respectively, the family  $\{\xi_x + \eta_x, \, x\in\Pi\BLam_L(\uu)\}$ is conditionally independent, given $\cB(\Lam_N(v))$, and so are the values $\{v(T^x\om,\eta), \, x\in\Pi \BLam_L(\uu)\}$.

Next, we re-write (4.6) as follows:
$$
\begin{array}{ll}
v(T^x\om;\th)  &=  \xi_x + a_{n_0(N)} \,  \th'_{n_0(N)} + \eta'_x,
\end{array}
\eqno(4.7)
$$
with
$$
\th'_{n_0(N)} = \th_{n_0(N),k(T^x\om)} \ffi_{n_0(N),k(T^x\om)}(T^x\om)
$$
and
$$
\eta'_x = \sum_{n > n_0(N)}^\infty  a_n \,  \th_{n,k(T^x\om)} \ffi_{n,k(T^x\om)}(T^x\om).
$$
The RV $\th'_{n_0(N)}$ is uniformly distributed in $[-a_{n_0(N)}, a_{n_0(N)}]$, so its probability density $p_{\th'_{n_0(N)}}$ exists and is bounded by $(2 a_{n_0(N)})^{-1}$. By virtue of (1.4) and (4.2), we have, therefore,
$$
a_{n_0(N)}^{-1} \le Const \, n_0(N)^M \le Const \ln^M N.
$$
The random variable $\eta'_x$ admits some probability density $p_{\eta'_x}$ (as a sum of a convergent series of
RVs with uniform distributions). Since $\th'_{n_0(N)}$ and $\eta'_x$ are independent,  their sum $\th'_{n_0(N)} + \eta'_x$ admits a probability density given by a convolution $p_{\th'_{n_0(N)}} * p_{\eta'_x}$, which is bounded by the $L^\infty$-norm of any of them. Hence,
$$
\|p\|_\infty \le (2 a_{n_0(N)})^{-1} \le Const \, \ln^{M} L.
$$

\subsection{Two-particle Wegner-type bound for independent potentials}\label{SubSectWeg2pIndep}

Here we recall the mains results of ref. \cite{CS08}, with necessary notational adaptations. Their proofs are based on the observation that finite-volume two-particle Hamiltonians with an external random potential can be represented as diagonally monotone operator families, so that Stollmann's method applies to such ensembles.

As before, $F_V$ is the marginal CDF of the external random potential.

\begin{proposition}[Cf. Thm. 1 in \cite{CS08}]\label{ThmW1} Consider a two-particle Hamiltonian $\uH = \uH_0 + (V(x_1;\om) + V(x_2;\om)) + U(x_1,x_2)$, where $\{V(x;\om), x\in\Pi\BLam_L(\uu)\}$ is an IID random field relative to a probability space
$(\Om, \cF, \P)$, with  a  marginal CDF $F_V(t)$. Set
$$
s(\eps) = \sup_{a\in\R} \;\;\int_a^{a+\eps} \, dF_V(t), \; 0 < \eps \le 1.
\eqno(4.7)
$$
Then for all $E\in\R$, $L\ge 1$, $\uu\in\Z^d\times\Z^d$
and $\epsilon >0$,
$$
\pr{\dist\left[\Sigma_{L,\uu}(\om;\th), E\right] \leq \eps }
\leq \left|\Lam_L(\uu )\right|^{3/2}\;  \cdot s(2\eps).
\eqno(4.8)
$$
\end{proposition}

\begin{proposition}[Cf. Thm. 2 in \cite{CS08}] Under the same assumptions as in Theorem \ref{ThmW1},
consider a pair of two-particle cubes $\BLam_L(\uu )$, $\BLam_{L'}(\uu' )$, $L \ge L'\ge 1$.
Let $S:(u_1,u_2)\mapsto (u_2,,u_1)$ be the symmetry in $\Z^{2d}$ exchanging the coordinates of two particles
and suppose that
$$
 \min \{ \|\uu -\uu'\|, \|\uu -S(\uu')\| \} > 8 L
 \eqno(4.9)
$$
Then for all $E\in\R$ and $\epsilon >0$,
$$
\begin{array}{l}
\pr{ \dist\left[\Sigma_{\uu,L}(\om;\th), \Sigma_{\uu',L'}(\om;\th) \right] \le \eps } \\
\leq |\BLam_L(\uu)|^{3/2} \cdot |\BLam_L(\uu')| \cdot s(2\eps).
\end{array}
\eqno(4.10)
$$
\end{proposition}

\subsection{One-volume Wegner-type bound for QP potentials}\label{SubSectQPW1}

Now we return to the analysis of our two-particle Hamiltonian $\uH_{\BLam_L(\uu)}$ with quasi-periodic potential $V(x;\om;\th)$ in a finite volume $\BLam_L(\uu)\subset\Z^{2d}$, where
$L\ge 1$, $\uu = (u_1, u_2)\in\Z^{2d}$,  $\BLam_L(\uu) = \Lam_L(u_1)\times \Lam_L(u_2)$, and $\Pi\BLam_L(\uu) \subset \Lam_N(v)$. Fix a real number $E$. We have to estimate the probability
$$
\pr{ \dist\left( \Sigma\left(\uH_{\BLam_L(\uu)}\right), \, E\right) \le \eps}.
\eqno(4.7)
$$
First, we can apply the identity
$$
\begin{array}{ll}
\pr{ \dist\left( \Sigma\left(\uH_{\BLam_L(\uu)}\right), \, E\right) \le \eps} \\
= \esm{ \pr{ \dist\left( \Sigma\left(\uH_{\BLam_L(\uu)}\right), \, E\right) \le \eps\, \big| \, \cB(\Lam_N(v))} },
\end{array}
\eqno(4.8)
$$
so that it suffices to bound the inner, conditional probability. It is clear that the results of the Subsection
\ref{SubSectWeg2pIndep}  apply to the two-particle Hamiltonian $\uH_{\BLam_L(\uu)}$ under conditioning by
the sigma-algebra $\cB(\Lam_N(v))$. Indeed, this conditioning makes the values of the external potential \textit{conditionally independent}. Moreover, the conditional probability density for each value of the external potential $V(x;\om;\th)$, $x\in\Lam_L(u_1) \cup \Lam_L(u_2)$, admits a density bounded by $a_{n_0(N)}^{-1}$
with $n_0(N) \le C' \ln N$. By assumption (1.4), such a density is bounded by
$C'' \, \ln^M N$, $C'',M<\infty$. Therefore, owing to (1.4), the following inequality holds true:
$$
 \pr{ \dist\left( \Sigma\left(\uH_{\BLam_L(\uu)}\right), \, E\right) \le \eps\, \big| \,\cB(\Lam_N(v))}
 \le C''' \ln^M N \cdot L^{3d}\cdot \eps.
$$
Finally, we see that, by virtue of (4.8),
$$
\pr{ \dist\left( \Sigma\left(\uH_{\BLam_L(\uu)}\right), \, E\right) \le \eps} \le C''' \ln^M N \cdot L^{3d}\cdot \eps.
$$
This concludes the proof of Theorem \ref{ThmMain1}. \QED

\subsection{Two-volume Wegner-type bound  for QP potentials}\label{SubSectQPW2}

We can argue as inthe one-volume case, but apply now Proposition 2 instead of Proposition 1. We deal here with a pair of two-particle cubes $\BLam_L(\uu )$, $\BLam_{L'}(\uu' )$, $L \ge L'\ge 1$. Suppose that $\Pi(\BLam_L(\uu') \cup \BLam_L(\uu'')) \subset \Lam_N(v) \subset \Z^d$ for some $N \in \db{N} , v \in \Z^d$ and condition(4.9) holds.
We have to estimate the probability
$$
\pr{ \dist\left[\Sigma_{\uu,L}(\om;\th), \Sigma_{\uu',L'}(\om;\th) \right] \le \eps }
$$
First, We notice that
$$
\begin{array}{ll}
\pr{ \dist\left[\Sigma_{\uu,L}(\om;\th), \Sigma_{\uu',L'}(\om;\th) \right] \le \eps } \\ \\
= \esm{\pr{ \dist\left[\Sigma_{\uu,L}(\om;\th), \Sigma_{\uu',L'}(\om;\th) \right] \le \eps
, \big| \, \cB(\Lam_N(v))}}.
\end{array}
$$
Therefore, one can apply the Proposition 2 of Subsection 4.3 to bound the conditional probability. Under this conditioning, the values of the external potential are independent, and the marginal conditional probability density for each value of the external potential $V(x;\om;\th)$, $x\in \Pi(\BLam_L(\uu') \cup \BLam_L(\uu''))$ is bounded by $a_{n_0(N)}^{-1} \le  C'' \, \ln^M N$.

Therefore the following inequality holds true:
$$
\pr{ \dist\left[\Sigma_{\uu,L}(\om;\th), \Sigma_{\uu',L'}(\om;\th) \right] \le \eps , \big| \, \cB(\Lam_N(v))}
 \le C''' L^{5d} \ln^M N \cdot \eps.
$$
This concludes the proof of Theorem \ref{ThmMain2}. \QED

%Author, Journal \textbf{Volume}, (year) page numbers.
%Author, \textit{Book title} (Publisher, place year) page numbers


\begin{thebibliography}{20}

\bibitem[BCKP88]{BCKP88} A. Bovier, M. Campanino, A. Klein, F. Perez, \textit{Smoothness of the density
of states in the Anderson model at high disorder}. - Comm. Math. Phys. (1988), {\bf 114}, 439-461.

\bibitem[BG00]{BG00} J. Bourgain, M. Goldstein, On nonperturbative localization with
quasiperiodic potentials. - Annals of  Math. (2000), {\bf 152}:3, 835-879.

\bibitem[BGS01]{BGS01} J. Bourgain, M. Goldstein, W. Schlag, Anderson localization for Schr\"{o}dinger
operators on $\Z$ with potential generated by skew-shift. - Comm. Math. Phys. (2001),
{\bf 220}, 583-621.

\bibitem[BS00]{BS00} J. Bourgain,  W. Schlag, Anderson localization for Schr\"{o}dinger
operators on $\Z$ with strongly mixing potential. - Comm. Math. Phys.
(2000), {\bf 215}, 143-175.

\bibitem[CamK86]{CamK86} M. Campanino, A. Klein,  \textit{A supersymmetric transfer matrix and differentiability
of the density of states in the one-dimensional Anderson model}. - Comm. Math. Phys.
(1986), {\bf 104}, 227-241.

\bibitem[CarL90]{CarL90} R. Carmona, J. Lacroix, \textit{Spectral Theory of Random Schr\"{o}dinger Operators}. -
Birkh\"{a}user, 1990.


\bibitem[C07]{C07} V. Chulaevsky, \textit{Wegner-Stollmann estimates for some quantum lattice systems.}
 - In: Contemporary Mathematics, v. 447, 17--89, 2007.


\bibitem[CS08]{CS08} V. Chulaevsky, Y. Suhov, \textit{ Wegner bounds for a two-particle tight binding model.} - Comm.
Math. Phys. (2008), {\bf 283}, 479-489.


\bibitem[Chan07]{Chan07} J. Chan, \textit{Method of variations of potential of quasi-periodic Schr\"{o}dinger
equations}. - Geom. Funct. Anal., {\bf 17}, 1416-1478, 2007.


\bibitem[CoHK07]{CoHK07} J.-M. Combes, P. D. Hislop, F. Klopp, \textit{An optimal Wegner estimate and
its application to the global continuity of the integrated density of states for
random Schr\"{o}dinger operators}. - Duke Math. J. (2007), {\bf 140}, no. 3, 469-498.


\bibitem[ConFS83]{ConFS83} F. Constantinescu, J. Fr\"{o}hlich, T. Spencer, \textit{Analyticity of the density
of states and replica method for random Schr\"{o}dinger operators on a lattice}. - J.
Statist. Phys. (1983), {\bf 34}, 571-596.

\bibitem[CyFKS87]{CyFKS87} H. L. Cycon, R. G. Froese, W. Kirsh, B. Simon, \textit{Shr\"{o}dinger
Operators}. - Springer-Verlag, 1987.


\bibitem[CrSim83]{CrSim83} W. Craig, B. Simon, \textit{Log H\"{o}lder continuity of the integrated density of
states for stochastic Jacobi matrices}. - Comm. Math. Phys.  (1983), {\bf 90}, 207-218.

\bibitem[DS84]{DS84} F. Delyon, B. Souillard, \textit{Remark on the continuity of the density of states
of ergodic finite difference operators}. -  Comm. Math. Phys.  (1984), {\bf 94}, 289.

\bibitem[DK89]{DK89} H. von Dreifus, A. Klein, \textit{A new proof of localization in the Anderson tight
binding model}. - Comm. Math. Phys. (1989), {\bf 124}, 285-299.

\bibitem[FS83]{FS83} J. Fr\"{o}hlich, T. Spencer, \textit{Absence of diffusion in the Anderson tight binding model for
large disorder or low energy}. - Comm. Math. Phys. {\bf 88}, 151-184, 1983.

\bibitem[FMSS85]{FMSS85} J. Fr\"{o}hlich, F. Martinelli, E. Scoppola, T. Spencer, \textit{A constructive proof of
localization in Anderson tight binding model}. - Comm. Math. Phys. (1985), {\bf 101},
21-46.

\bibitem[FSW87]{FSW87} J. Fr\"{o}hlich, T. Spencer, P. Wittwer, \textit{Localization for a Class of One
Dimensional Quasi-Periodic Schr\"{o}dinger Operators}. - Preprint, 1987; Comm. Math. Phys.
{\bf 1332} (1990), 5-26.


\bibitem[PF92]{PF92} L. A. Pastur, A. L. Figotin, Spectra of Random and Almost Periodic
Operators, Springer-Verlag, Berlin, 1992.

\bibitem[ST85]{ST85} B. Simon , M. Taylor, \textit{Harmonic analysis on $SL(2,\R)$ and smoothness
of the density of states in the Anderson model}. - Comm. Math. Phys. (1985), {\bf 101},
1-19.


\bibitem [St00] {St00}  P. Stollmann:  \textit{Wegner estimates and localization for continuous
Anderson models with some singular distributions}. - Arch. Math., \textbf{ 75}, 307--311, 2000.

\bibitem[St01]{St01} P. Stollmann, \textit{Caught by Disorder. A Course on Bound States in Random Media.} -
Birkh\"{a}user (2001).

\bibitem[W81]{W81} F. Wegner, \textit{Bounds on the density of states in disordered systems}. - Z.
Phys. {bf B}. Condensed Matter  (1981), {\bf 44}, 9-15.


\end{thebibliography}
\end{document}